\documentclass[runningheads]{llncs}
\usepackage[misc]{ifsym}
\usepackage{times}

\usepackage{soul}
\usepackage{url}
\usepackage[hidelinks]{hyperref}
\usepackage[utf8]{inputenc}
\usepackage[small]{caption}
\usepackage{graphicx}
\usepackage{amsmath}
\usepackage{booktabs}
\usepackage{mathrsfs}
\usepackage{cite}

\usepackage{color}
\usepackage{amsfonts,amssymb}
\begin{document}
\title{Uncertainty-Guided Lung Nodule Segmentation with Feature-Aware Attention\thanks{Corresponding author: Q.~Wang. $^\dag$ indicates equal contribution.}}

\titlerunning{ }

\author{Han~Yang$^\dag$\inst{2} \and
Lu~Shen$^\dag$\inst{2} \and
Mengke~Zhang\inst{2}\and
Qiuli~Wang\inst{1}}
\authorrunning{H. Yang, L. Shen, et al.}
%
\institute{
  School of Biomedical Engineering and Suzhou Institute for Advanced Research, University of Science and Technology of China, Suzhou, China\\
  \email{wangqiuli@ustc.edu.cn}\\ \and
  School of Big Data and Software Engineering, Chongqing University, Chongqing, China\\
\email{\{yang\_han,shenlu,mkzhang\}@cqu.edu.cn}}

\maketitle              
\begin{abstract}
Since radiologists have different training and clinical experiences, they may provide various segmentation annotations for a lung nodule. 
Conventional studies choose a single annotation as the learning target by default, but they waste valuable information of consensus or disagreements ingrained in the multiple annotations. 
This paper proposes an Uncertainty-Guided Segmentation Network (UGS-Net), which learns the rich visual features from the regions that may cause segmentation uncertainty and contributes to a better segmentation result. 
With an Uncertainty-Aware Module, this network can provide a Multi-Confidence Mask (MCM), pointing out regions with different segmentation uncertainty levels. 
Moreover, this paper introduces a Feature-Aware Attention Module to enhance the learning of the nodule boundary and density differences. 
Experimental results show that our method can predict the nodule regions with different uncertainty levels and achieve superior performance in LIDC-IDRI dataset. 
\keywords{Lung Nodule \and Segmentation \and Uncertainty \and Attention Mechanism \and Computed Tomography.}
\end{abstract}

\section{Introduction}
Lung nodule segmentation plays a crucial role in Computer-Aided Diagnosis (CAD) systems for lung nodules \cite{liu2019cascaded}.
Traditional methods like \cite{zhu2018deeplung,xie2019automated,gonccalves2016hessian,wu2010stratified,pezzano2021cole} usually use a single annotation as the learning target, which is consistent with conventional deep learning strategy. 
However, in the clinical situation, a sample lung nodule may be evaluated by several radiologists, and datasets like LIDC-IDRI \cite{armato2011lung} also provide multiple annotations for a lung nodule. As a result, these methods cannot make the most of the information in multiple annotations.

The challenge of learning multiple annotations is that there might be uncertain regions, which are annotated as `nodule tissues' by some doctors and annotated as `not nodule tissues' by the other. 
To this end, studies like \cite{hu2019supervised,kohl2019hierarchical,xiaojiang2021} propose to model these uncertainties as probabilistic distributions and produce segmentation masks with the random variables in latent space. 
Simon A. A. Kohl \cite{kohl2019hierarchical} proposed a model based on probabilistic U-Net \cite{kohl2018probabilistic} with a conditional variational auto-encoder that used a hierarchical latent space decomposition.
X.~Long \cite{xiaojiang2021} proposed to learn the probabilistic distributions of multiple 3D annotations for a lung nodule with a probabilistic V-Net.
But these methods cannot have stable segmentation results relying on random variables. 

\begin{figure}[t]
  \centerline{\includegraphics[width=100mm]{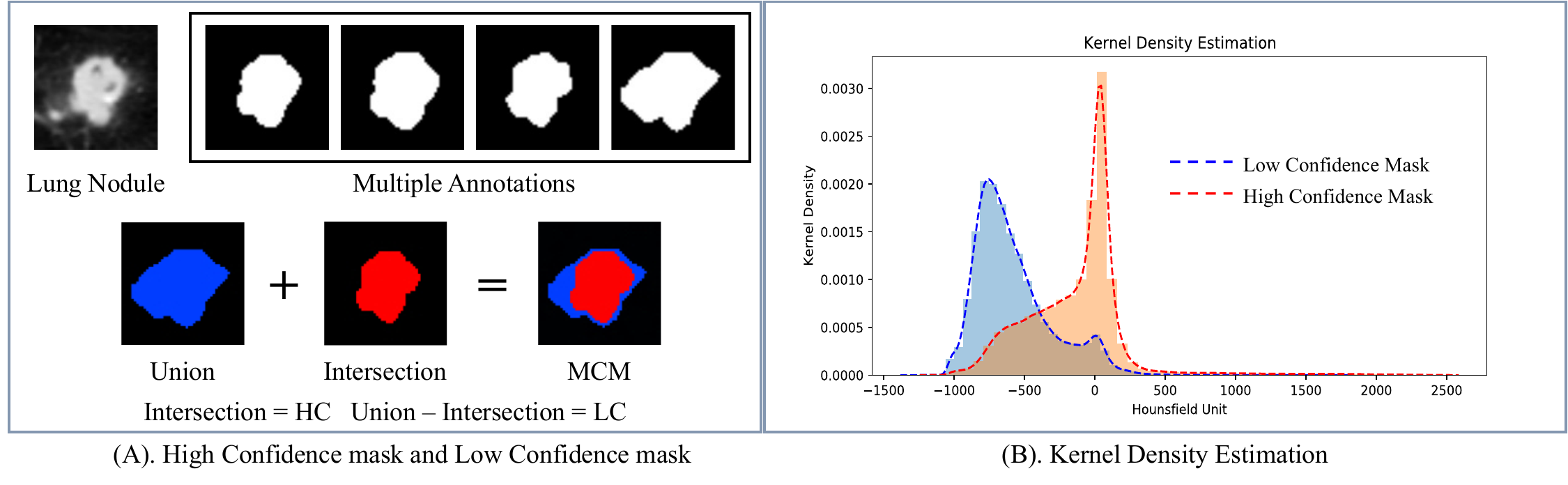}}
  \vspace{-0.0cm}
  \caption{
    \textbf{(A)}: Overview of High Certainty (HC) mask, Low Certainty (LC) mask, and Multi-Confidence Mask (MCM). 
    \textbf{(B)}: Kernel density estimation of HC and LC masks in LIDC-IDRI. The HU values in LC are mainly distributed around \emph{HU -750}, and the HU values in HC are mainly distributed around \emph{HU 0}.
  }
  \vspace{-0.0cm}
  \label{dif}
  \end{figure} 

This paper proposes that regions causing segmentation uncertainty are not \emph{random} and introduces a Multi-Confidence Mask (MCM).
As shown in Figure~\ref{dif}(A), the MCM defines two categories for the mask pixels: High Confidence (HC) mask and Low Confidence (LC) mask. The HC mask indicates the regions annotated by all radiologists (intersection of all annotations), and the LC mask shows the regions that the radiologists have disagreements (the difference of union and intersection). \emph{The question is, what's the difference between the HC and LC masks?}

To demonstrate the difference between HC and LC masks, we further calculate the HU (Hounsfield Unit) Kernel Estimations in HC and LC of LIDC-IDRI \cite{armato2011lung}. As shown in Figure~\ref{dif}(B), HC and LC masks have quite different HU distributions: HU values in HC masks are mainly distributed around \emph{HU 0}, while HU values in LC masks are mainly distributed around \emph{HU -750}. 
HU reflects tissues' density \cite{gao2018holistic}. 
Additionally, we can find that LC is basically around the edge of nodules from Figure~\ref{dif}(A).
These phenomenons indicate that the radiologists may have more diverse opinions about low-dense nodule tissues and boundaries.

Therefore, our study focuses on two challenges: (1) Making good use of all annotations during model training. (2) Utilizing radiologists' consensus and disagreements to improve segmentation performance.
To tackle these challenges, we propose to learn the consensus and disagreements in all annotations directly without changing the end-to-end training strategy. 
Our contributions can be summarized as: 
(1) To enable the network to learn the information of multiple annotations, this paper introduces a Multi-Confidence Mask (MCM) composed of all annotations' union and intersection, reflecting lung nodule regions with different uncertainty levels, which can be used as references for clinical use. 
(2) This paper proposes an Uncertainty-Guided Segmentation Network (UGS-Net), which contains an Uncertainty-Aware Module (UAM) and a Feature-Aware Attention Module (FAAM). The UAM introduces the consensus and disagreements in the annotation sets into the learning process. At the same time, the FAAM guides the network to learn more specific features about ambiguous regions, like boundaries and low-dense tissues, further improving the segmentation performance.

\section{Methods}
\label{method}
\begin{figure}[t]
\centerline{\includegraphics[width=110mm]{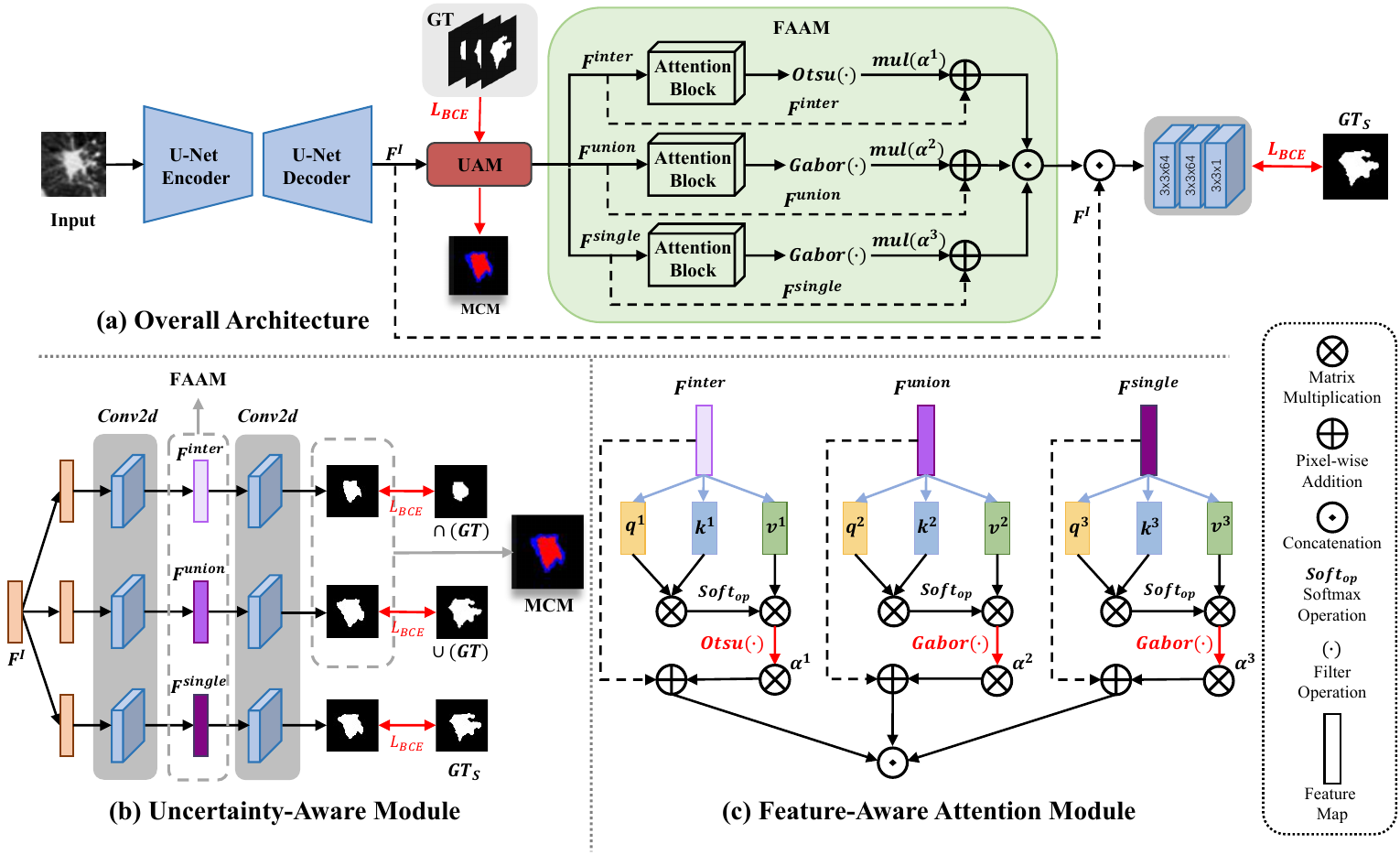}}
\vspace{-0.0cm}
\caption{
  \textbf{(a)}: Overview of Uncertainty-Guided Segmentation Network.
  \textbf{(b)}: Uncertainty-Aware Module. It takes the feature maps $F^I$ as the input, and then fuses them into three branches with three learning targets: $\cup(GT)$, $\cap(GT)$, and $GT_{S}$. 
  \textbf{(c)}: Feature-Aware Attention Module.
}
\vspace{-0.0cm}
\label{arch}
\end{figure}

Figure~\ref{arch} shows the architecture of the proposed Uncertainty-Guided Segmentation Network (UGS-Net). 
UGS-Net's input is the lung nodule CT image $X$, and its final learning target keeps consistent with the current mainstream methods, which is a single annotation $GT_{S}$. 
Moreover, we introduce two additional auxiliary learning targets, multiple annotation's union $\cup(GT)$ and intersection $\cap(GT)$, to assist the proposed network in learning richer information from disagreements among multiple annotations. 
The input images and their masks have the same size $64\times64$.

The UGS-Net contains a coarse to fine three-stage processing pipeline:
\textbf{(1)} UGS-Net uses an Attention U-Net \cite{oktay2018attention} to capture deep feature maps $F^I$ with 64 channels. This Attention U-Net has five down-sampling and up-sampling layers. Each up-sampling layer is composed of two convolutional layers and an attention block.
\textbf{(2)} the Uncertainty-Aware Module (UAM) analyzes the $F^I$ and generates $\cup(GT)'$, $\cap(GT)'$, and $GT_{S}'$ under the guidance of $\cup(GT)$, $\cap(GT)$, and $GT_{S}$. Additionally, three visual feature variables $F^{union}$, $F^{inter}$, and $F^{single}$ are obtained in UAM and will be fed into the next module. UAM can generate MCM simultaneously, pointing out lung nodule regions with different uncertainty levels.
\textbf{(3)} the Feature-Aware Attention Module (FAAM) further analyzes the $F^{union}$, $F^{inter}$, and $F^{single}$ with different learning preferences by introducing the visual filters to the self-attention mechanism.
The FAAM output is concatenated with $F^I$. Then they will be fed into the output CNN module, which contains three convolutional layers. The output of this module is the final segmentation result $S$.

\subsection{Uncertainty-Aware Module}
The consensus and disagreements among multiple experts reflect the uncertainty levels of different nodule tissues in CT images.
How to better take advantage of this valuable uncertainty information to further improve the overall segmentation performance is an important research problem.
In this paper, we propose an Uncertainty-Aware Module (UAM) and introduce two auxiliary learning targets to take full utilization of all the annotation information.

Specifically, we define a \emph{Union Mask} $\cup(GT)$ and an \emph{Intersection Mask} $\cap(GT)$ as two auxiliary learning targets besides the final learning target \emph{Single Mask} $GT_{S}$, where $GT$ is a collection of all the annotations.
As shown in Figure~\ref{arch}(b), the backbone of UAM is a three-branch CNN network and each branch has two convolutional layers.
This module takes the deep feature maps $F^I$ with 64 channels as the input and its loss is computed as $L_{UAM} = L_{BCE}(\cup(GT)', \cup(GT)) + L_{BCE}(\cap(GT)',\cap(GT)) + L_{BCE}(GT_{S}', GT_{S})$,
where the $\cup(GT)'$, $\cap(GT)'$, and $GT_{S}'$ represent the results of the three branches. $L_{BCE}$ denotes the binary cross entropy loss.

The outputs of UMA contain two parts. 
The first part is Multi-Confidence Mask defined as $normalization(\cup(GT)' + \cap(GT)')$, which can show different uncertainty levels in different nodule regions.
Moreover, under the guidance of different learning targets, three feature maps $F^{union}$, $F^{inter}$, and $F^{single}$ with different latent visual features will be obtained after the first convolutional layer. 
These three feature maps are the second part of outputs, and they will be fed into the next module for further analysis. 

\subsection{Feature-Aware Attention Module}
To further distinguish three feature maps ($F^{union}$, $F^{inter}$, and $F^{single}$) and extract more features with high uncertainty, a Feature-Aware Attention Module (FAAM) is introduced to capture more density differences and boundaries of nodules.

As shown in Figure~\ref{arch}(c), the Feature-Aware Attention Module (FAAM) has three attention blocks, and each block contains a self-attention block \cite{vaswani2017attention} and a feature-aware filter. These attention blocks process $F^{union}$, $F^{inter}$, and $F^{single}$ with different feature-aware filters, which enable the network to formulate different learning preferences for different learning objectives, to obtain more image features that are helpful for segmentation.
More specifically, assuming the input is $F^z$, for each attention block, we have $o_z=\Gamma(A(F^z))$, where $z\in\{union,inter,single\}$, $A$ indicates the self-attention architecture, and $\Gamma$ is a feature-aware filter. The process of $A$ please refer to the study \cite{vaswani2017attention}. The output of FAAM is $\odot\{o_{union}, o_{inter}, o_{single}\}$, where $\odot$ is the concatenation operation. Each $o_z$ has 32 channels, so $\odot\{o_{union}, o_{inter}, o_{single}\}$ has 96 channels in total. 

As mentioned above, each learning objective has its own filter setting. 
We use Otsu \cite{2007A} as its feature-aware filter when learning the $F^{inter}$, because Otsu is more sensitive to high dense nodule tissues \cite{wang2021realistic}.
We choose Gabor as their feature-aware filter when learning $F^{union}$ and $F^{single}$ for two reasons.
Firstly, Gabor filter is sensitive to image edges and can provide good directional selection and scale selection features, so that it can better capture the nodule boundary features in $F^{union}$. 
Secondly, the Gabor filter can capture the local structural features of multiple directions in the image's local area, which is very suitable for extracting the overall general features in $F^{single}$.
To improve the calculation speed, we use GConv in Gabor CNN \cite{luan2018Gabor}, which can also enhance the adaptability of deep learning features to changes in orientation and scale.

Finally, the output CNN module that mentioned above processes the concatenation of $\odot\{o_{union}, o_{inter}, o_{single}\}$ and $F^{I}$, and the final segmentation result $S$ is generated at the end of this module.
The total training loss $L$ for the proposed UGS-Net can be represented as $L = L_{BCE}(S, GT_{S}) + L_{UAM}$.    

\section{Experiments and Results}
\label{experiments}
\subsection{Dataset and Experimental Setup}
This study uses LIDC-IDRI \cite{armato2011lung} dataset, which has 2635 annotated lung nodules with malignancy labels. 
We extract 1859 nodules with multiple annotations (more than 1) and their annotations' union and intersection.
We use five-cross validation in this study and ensure each fold has the same malignant/benign nodule distributions.
We use the Stochastic Gradient Descent with Warm Restarts (SGDR) \cite{2016SGDR} to optimize parameters and build the network with PyTorch. All experiments are run on a GPU of NVIDIA Tesla V100. The UGS-Net needs 23 GB GPU memory, and is trained with 150 epochs. The batch size is set as 32. More details can be found in supplementary materials. We will release the source code for both data pre-processing and the proposed network.

This paper uses Dice, IoU, and NSD as the evaluation metrics. 
For the convenience of explanation, we temporarily define the ground truth as $G$ and segmentation result as $S$.
Dice is used to calculate the similarity of $G$ and $S$. IoU is the ratio of $G$ and $S$ \cite{diceiouweichen,8576421}. 
Normalized surface Dice (NSD) \cite{ma2021abdomenct} is used to evaluate how close $G$ and $S$ are to each other at a specified tolerance $\tau$. The NSD is defined as:
\begin{equation}
  NSD(G,S)=\frac{\vert\partial G \cap B_{\partial S}^{\left(\tau \right)} \vert + \vert\partial S \cap B_{\partial G}^{\left(\tau \right)} \vert}{\vert\partial G\vert + \vert\partial S\vert}    
\end{equation}
where $\vert\partial G\vert$ and $|\partial S|$ are the number of pixels of G and S. $B_{\partial G}^{\left(\tau \right)}$, $B_{\partial S}^{\left(\tau \right)} \subset R^{3}$ and $B_{\partial G}^{\left(\tau \right)} = \left\{x \in R^{3} | \exists \tilde{x} \in \partial G, ||x - \tilde{x}|| \leq \tau\right\}$, $B_{\partial S}^{\left(\tau \right)} = \left\{x \in R^{3} | \exists \tilde{x} \in \partial S, ||x - \tilde{x}|| \leq \tau\right\}$. 

\subsection{Performance of General Segmentation}
\subsubsection{Lung Nodule Segmentation}
\begin{figure}[t]
\centerline{\includegraphics[width=80mm]{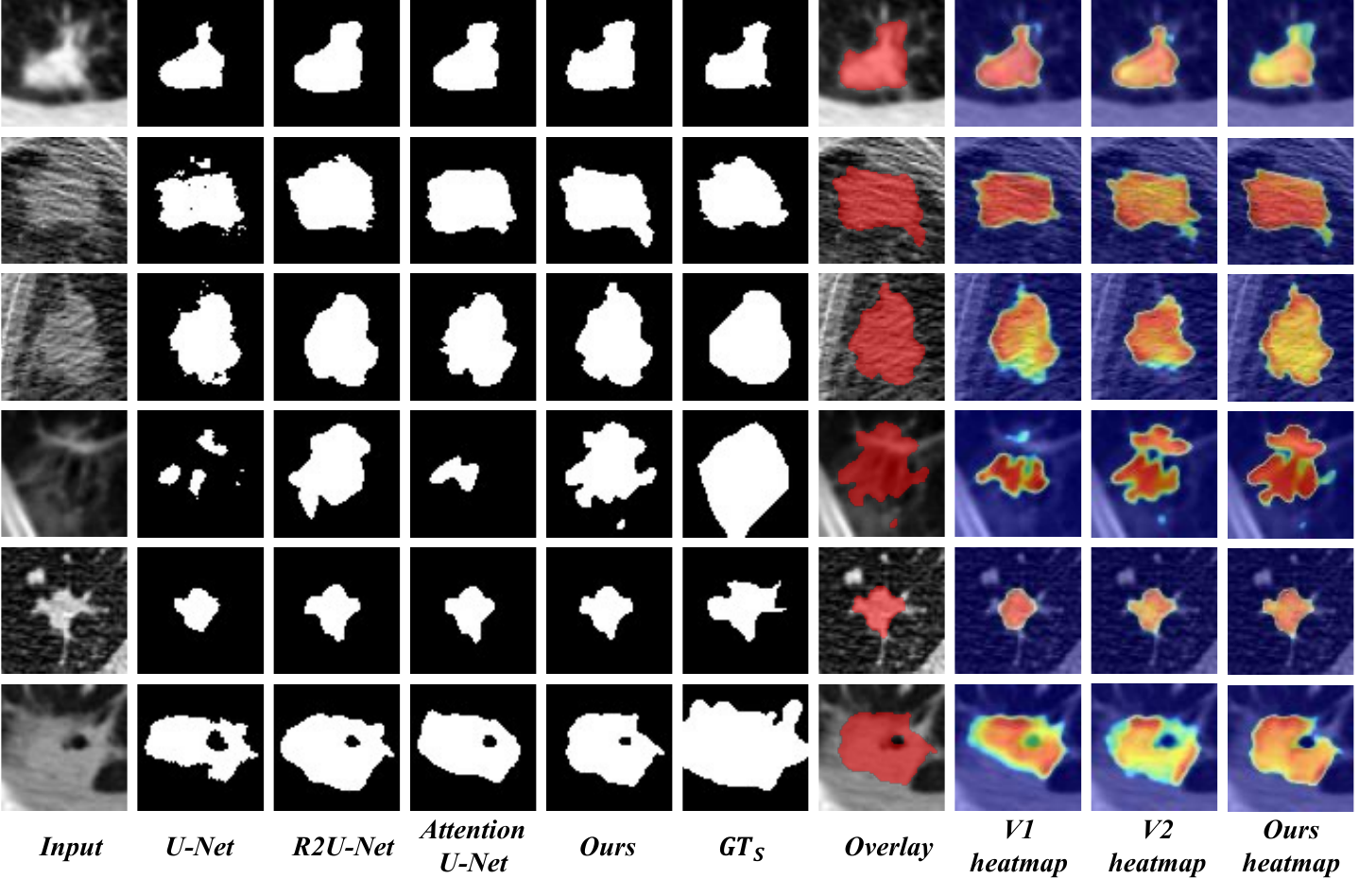}}
\vspace{-0.0cm}
\caption{Segmentation Results of Different Methods.}
\vspace{-0.0cm}
\label{compare}
\end{figure}

We compare our network with the state-of-the-art (SOTA) methods for medical image segmentation which are trained with single annotation. 
Table~\ref{vsother} quantitatively compares UGS-Net with nine SOTA lung nodule segmentation methods.
\begin{table}[h]
  \vspace{-0.0cm}
  \caption{Comparison between our framework and existing methods on the LIDC–IDRI dataset.}  
  \begin{center}{
  \begin{tabular}{c|c|c|c}
  \hline
  \textbf{\emph{Method}} & \textbf{\emph{Dice($\%$)}} & \textbf{\emph{IoU($\%$)}} & \textbf{\emph{NSD($\%$)}} \\
  \hline
  FCN V-Net \cite{liu2019cascaded} & 79.59 &  - & - \\
  Dual-Branch ResNet \cite{cao2020dual}& 82.74 &  - & - \\
  nnU-Net \cite{isensee2021nnu} & 85.82 &  - & - \\
  R2U-Net \cite{alom2018recurrent}& 82.10 & 71.18 & 93.07 \\
  Attention U-Net \cite{oktay2018attention}\ & 85.37 & 75.45 & 94.80 \\
  Multi-Orientation U-Net \cite{amorim2019lung}& 83.00 & 76.00 & - \\
  3D DCNN \cite{tang2019nodulenet}& 83.10 & 71.85 & - \\
  Channel U-Net \cite{tolooshams2020channel}& 84.02 & 73.53  & - \\
  Nested U-Net \cite{zhou2018unet++}& 83.44 & 72.72 & - \\
  \hline
  Baseline (U-Net) & 85.05 & 75.27 & 94.43 \\
  Attention U-Net+UAM (V1)& 85.94 & 76.20 & 95.06 \\
  V1+self-attention block (V2)& 85.97 & 76.18 & 95.22 \\
  UGS-Net & \textbf{86.12}& \textbf{76.44} & \textbf{95.36} \\
  \hline
  \end{tabular}}  
  \label{vsother}
  \end{center}   
  \vspace{-0.0cm}
\end{table}
According to Table~\ref{vsother}, the average Dice score of our network is 86.12$\%$, which achieves superior performance in all compared methods. 
At the same time, our average IoU and NSD are 76.44$\%$ and 95.36$\%$, which are also better than other methods. 
The first seven columns of Figure~\ref{compare} show the segmentation results of UGS-Net and other networks. 
It is obvious that the segmentation results generated by the proposed network are better than other methods, especially for the ambiguous regions among different experts, such as low-dense tissues, edge areas, and spiculations.
We choose Attention U-Net as our first component even though nnU-Net achieves better performance. That's because the nnU-Net needs over 150 GB of system memory and takes too much system resources, but the performance is slightly better than the Attention U-Net.

\subsubsection{Ablation Studies}
This section will evaluate the performance of each component in the UGS-Net.
Firstly, by integrating the UAM into Attention U-Net, the Dice score increases by 0.89$\%$.
It indicates that the calibration prediction ability of the model can be improved by introducing the consensus and disagreements of doctors' professional knowledge to guide model learning. 
Secondly, to better use the uncertain clues provided by doctors, the FAAM is designed to conduct targeted learning of nodules' different features under the guidance of these clues, which improves Dice, IoU, and NSD scores, respectively.
Compared with integrating the self-attention block (Table~\ref{vsother}(V2)) into the model directly, 
the FAAM (Table~\ref{vsother}(UGS-Net)) achieves better performance, which indicates filtering can enhance the differentiation of attention in different branches and enable the model to extract features more comprehensively.

To further explore the influence of UAM and FAAM, we visualize the network's last convolution layer in different configurations with Grad CAM\cite{2020Grad}, and show the results in Figure~\ref{compare}(column8-10). 
According to this figure :(1) our network can identify cavities more clearly (Figure~\ref{compare}(row6)); (2) our network can notice more low-density tissues (Figure~\ref{compare}(row1, 2, 4)); (3) our network is more sensitive to spiculation (Figure~\ref{compare}(row3, 5)); (4) for nodules with higher density, model's attention will gradually turn to the boundary(Figure~\ref{compare} (row3)). 
More results are provided in the supplementary materials.

\subsection{Performance of Multi-Confidence Mask}
\subsubsection{Prediction of Intersection and Union mask}
The quality of the predicted union and intersection directly affects the final estimation of Multi-Confidence Mask (MCM).
To demonstrate the UGS-Net's ability to predict union and intersection, we design a comparative experiment.
The experimental results are shown in Table \ref{table_MCM}.

According to Table~\ref{table_MCM}, the U-Net baseline achieves an average Dice score of 85.05$\%$ with \emph{Single Mask}. 
Meanwhile, it achieves average Dice scores of 84.08$\%$ and 82.66$\%$ with \emph{Intersection Mask} and \emph{Union Mask}. 
The proposed UGS-Net shows the optimal performance in most indicators. 
Compared with baseline, UGS-Net improves all indicators of \emph{Single mask} and \emph{Intersection mask}, in which Dice scores are improved by 1.07$\%$ and 0.63$\%$, IoU scores are improved by 1.17$\%$ and 0.55$\%$, and NSD scores are improved 0.83$\%$ and 0.35$\%$, respectively. 
This indicates that the proposed network enhances the recognition of most nodule tissues, and has a more accurate calibration of nodule boundaries. 
Additionally, UGS-Net makes greater progress in \emph{Union mask}, whose Dice score improved by 1.7$\%$ compared to the baseline, IoU improved by 2.03$\%$, and NSD is improved by 1.1$\%$. 
Experimental results show that with the help of UAM and FAAM, our network can effectively utilize the disagreements of doctors to learn more high uncertainty nodule regions.

\begin{table}[h]
  \vspace{-0.0cm}
  \caption{Performances of UGS-Net and other methods on the $\cup(GT)$ and $\cap(GT)$ segmentation.}  
  \begin{center}{
  \setlength{\tabcolsep}{1mm}{
  \begin{tabular}{c|c|c|c|c|c|c|c|c|c}
  \hline
  \textbf{\emph{Method}} & \multicolumn{3}{|c|}{\textbf{\emph{Single mask}}} & \multicolumn{3}{|c|}{\textbf{\emph{Intersection mask}}} & \multicolumn{3}{|c}{\textbf{\emph{Union mask}}} \\
  \hline
  & \textbf{\emph{Dice}} & \textbf{\emph{IoU}} & \textbf{\emph{NSD}} & \textbf{\emph{Dice}} & \textbf{\emph{IoU}} & \textbf{\emph{NSD}} & \textbf{\emph{Dice}} & \textbf{\emph{IoU}} & \textbf{\emph{NSD}} \\
  \hline
  U-Net \cite{ronneberger2015u} & 85.05 & 75.27 & 94.43 & 84.08 & 73.92 & 95.20 & 82.66 & 71.75 & 92.62 \\
  R2U-Net \cite{alom2018recurrent} & 82.10 & 71.18 & 93.07 & 77.90 & 65.70 & 92.06 & 83.12 & 72.28 & 92.71 \\
  Attention U-Net \cite{oktay2018attention} & 85.37 & 75.45 & 94.80 & 84.07 & 73.71 & 95.10 & 83.48 & \textbf{73.93} & 93.30 \\
  \hline
  UGS-Net & \textbf{86.12}& \textbf{76.44} & \textbf{95.26} & \textbf{84.71} & \textbf{74.47} & \textbf{95.55} & \textbf{84.36} & 73.78 & \textbf{93.72} \\
  \hline
  \end{tabular}}} 
  \label{table_MCM}
  \end{center} 
  \vspace{-0.0cm}
\end{table}

\subsubsection{Prediction of Multi-Confidence Mask}
\begin{figure}[h]
  \centerline{\includegraphics[width=120mm]{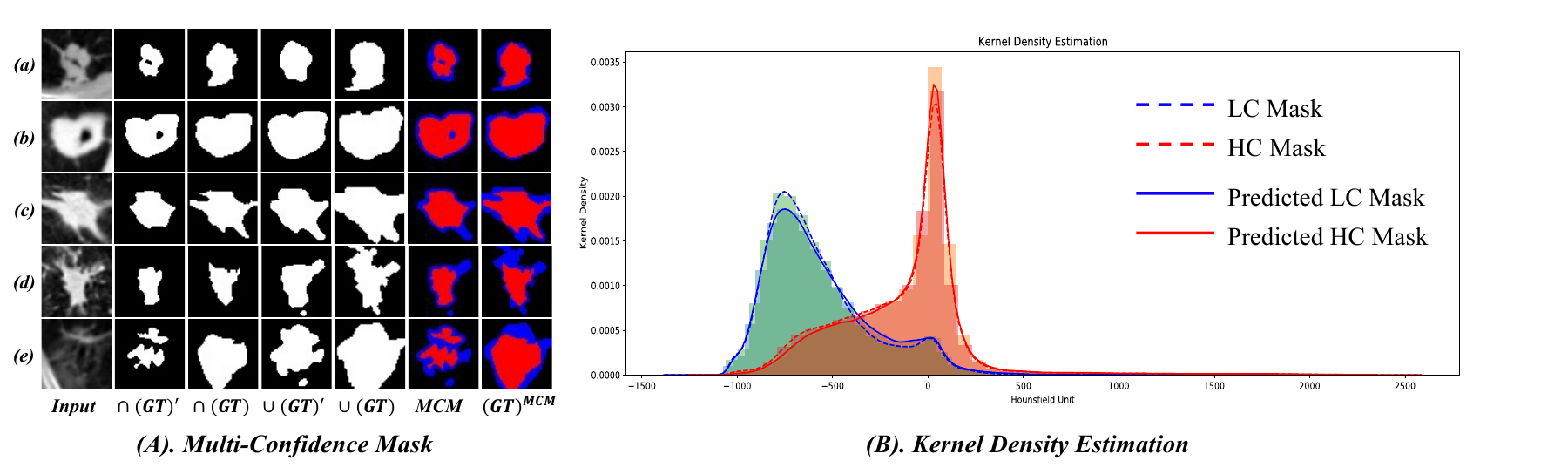}}
  \vspace{-0.0cm}
  \caption{
    \textbf{(A)}: Prediction of the Multi-Confidence Mask. For each nodule, we provide the predicted intersection ($\cap(GT)^{'}$), the standard intersection ($\cap(GT)$), the predicted union ($\cup(GT)^{'}$), the standard union ($\cup(GT)$), and the predicted MCM, the standard MCM ($(GT)^{MCM}$). 
    \textbf{(B)}: Kernel density estimation for the predicted HC and LC.
  }
  \vspace{-0.0cm}
  \label{MCM}
  \end{figure}

Figure~\ref{MCM}(A) shows the generated MCM and the prediction result of multiple annotations' intersection, union.
Colors are used to show the differences, and the original outputs are grayscale.
As can be seen, our MCM has several advantages:
(1) MCM can better show the significant cavity features of lung nodules (Figure~\ref{MCM}(A).(a)(b)). On the prediction of the intersection, the cavity is obviously larger, and we believe it is because our network can better capture the density differences about nodule tissues, and keep more features about cavity.
(2) MCM can better show the spiculation, which is an important feature for the diagnosis of the malignant nodule (Figure~\ref{MCM}(A).(c)(d)).
The spiculation is a stellate distortion caused by the intrusion of nodules into surrounding tissue, which is low-dense and distributed around the nodule edges. The network pays more attention to the low-dense tissues and boundaries when learning the union of multiple annotations. As a result, it naturally has better performance on spiculation segmentation.
(3) MCM can segment the part-solid nodule (Figure~\ref{MCM}(A).(e)) better, which means the network can take advantage of the multiple annotations since the experts tend to have different opinions in the low-dense regions.

Most evaluation metrics can not directly measure the quality of MCM prediction due to the small LC regions.
So we calculate the HU distributions in predicted HC and LC masks, and compare them with the real ones. 
If our network can predict the regions with different uncertainty levels well, the HU distributions of predicted HC and LC should be similar to the real one. 
As shown in this Figure~\ref{MCM}(B), the predicted and real curves are almost the same, which means our predicted uncertainty levels of regions are convincing from the perspective of statistics.

\section{Conclusions}
\label{conclusions}
This paper presents a novel Uncertainty-Guided framework for lung nodule segmentation, which focuses on utilizing richer uncertainty information from experts' diverse opinions and reflects better segmentation performance. 
We introduce an Uncertainty-Aware Module and a Feature-Aware Attention Module to fully learn the corresponding features under the guidance of three different learning targets and improve lung nodule segmentation performance.
Additionally, the network can generate a Multi-Confidence Mask to show nodule regions with different uncertainty levels.
Experimental results show that our method can predict lung nodule regions with high uncertainty levels and outperform many state-of-the-art networks.

\bibliographystyle{splncs04}
\bibliography{sigproc} 

\begin{thebibliography}{10}
\providecommand{\url}[1]{\texttt{#1}}
\providecommand{\urlprefix}{URL }
\providecommand{\doi}[1]{https://doi.org/#1}

\bibitem{liu2019cascaded}
Liu, H., Cao, H., Song, E., \emph{et al.}: A cascaded dual-pathway residual
  network for lung nodule segmentation in ct images. PHYS MEDICA  \textbf{63},
  112--121 (2019)

\bibitem{zhu2018deeplung}
Zhu, W., Liu, C., Fan, W., \emph{et al.}: Deeplung: Deep 3d dual path nets for
  automated pulmonary nodule detection and classification. In: Proc. IEEE WACV.
  pp. 673--681. IEEE (2018)

\bibitem{xie2019automated}
Xie, H., Yang, D., Sun, N., \emph{et al.}: Automated pulmonary nodule detection
  in ct images using deep convolutional neural networks. PATTERN RECOGN
  \textbf{85},  109--119 (2019)

\bibitem{gonccalves2016hessian}
Gonçalves, L., Novo, J., Campilho, A.: Hessian based approaches for 3d lung
  nodule segmentation. EXPERT SYST APPL  \textbf{61},  1--15 (2016)

\bibitem{wu2010stratified}
Wu, D., Lu, L., Bi, J., \emph{et al.}: Stratified learning of local anatomical
  context for lung nodules in ct images. In: Proc. IEEE CVPR. pp. 2791--2798
  (2010)

\bibitem{pezzano2021cole}
Pezzano, G., Ripoll, V.R., Radeva, P.: Cole-cnn: Context-learning convolutional
  neural network with adaptive loss function for lung nodule segmentation.
  COMPUT METH PROG BIO  \textbf{198},  105792 (2021)

\bibitem{armato2011lung}
Armato~III, S., McLennan, G., Bidaut, L., \emph{et al.}: The lung image
  database consortium (lidc) and image database resource initiative (idri): a
  completed reference database of lung nodules on ct scans. MED PHYS
  \textbf{38}(2),  915--931 (2011)

\bibitem{hu2019supervised}
Hu, S., Worrall, D., Knegt, S., \emph{et al.}: Supervised uncertainty
  quantification for segmentation with multiple annotations. In: Proc. MICCAI.
  pp. 137--145. Springer (2019)

\bibitem{kohl2019hierarchical}
Kohl, S.A., Romera-Paredes, B., Maier-Hein, K.H., \emph{et al.}: A hierarchical
  probabilistic u-net for modeling multi-scale ambiguities. arXiv preprint
  arXiv:1905.13077  (2019)

\bibitem{xiaojiang2021}
Long, X., Chen, W., Wang, Q., \emph{et al.}: A probabilistic model for
  segmentation of ambiguous 3d lung nodule. In: Proc. IEEE ICASSP. pp.
  1130--1134. {IEEE} (2021). \doi{10.1109/ICASSP39728.2021.9415006}

\bibitem{kohl2018probabilistic}
Kohl, S., Romera-Paredes, B., Meyer, C., \emph{et al.}: A probabilistic u-net
  for segmentation of ambiguous images. Proc. NIPS  \textbf{31} (2018)

\bibitem{gao2018holistic}
Gao, M., Bagci, U., Lu, L., \emph{et al.}: Holistic classification of ct
  attenuation patterns for interstitial lung diseases via deep convolutional
  neural networks. Comput Methods Biomech Biomed Eng Imaging Vis
  \textbf{6}(1), ~1--6 (2015)

\bibitem{oktay2018attention}
Oktay, O., Schlemper, J., Folgoc, L.L., \emph{et al.}: Attention u-net:
  Learning where to look for the pancreas. arXiv preprint arXiv:1804.03999
  (2018)

\bibitem{vaswani2017attention}
Vaswani, A., Shazeer, N., Parmar, N., \emph{et al.}: Attention is all you need.
  Proc. NIPS  \textbf{30} (2017)

\bibitem{2007A}
Otsu, N.: A threshold selection method from gray-level histograms. IEEE T SYST
  MAN CY-S  \textbf{9}(1),  62--66 (2007)

\bibitem{wang2021realistic}
Wang, Q., Zhang, X., Zhang, W., \emph{et al.}: Realistic lung nodule synthesis
  with multi-target co-guided adversarial mechanism. IEEE T MED IMAGING
  \textbf{40}(9),  2343--2353 (2021)

\bibitem{luan2018Gabor}
Luan, S., Chen, C., Zhang, B., Han, J., \emph{et al.}: Gabor convolutional
  networks. IEEE T IMAGE PROCESS  \textbf{27}(9),  4357--4366 (2018)

\bibitem{2016SGDR}
Loshchilov, I., Hutter, F.: Sgdr: Stochastic gradient descent with warm
  restarts  (2016)

\bibitem{diceiouweichen}
Chen, W., Wang, K., Yang, D., \emph{et al.}: {MTGAN:} mask and texture-driven
  generative adversarial network for lung nodule segmentation. In: Proc. IEEE
  ICPR. pp. 1029--1035. {IEEE} (2020). \doi{10.1109/ICPR48806.2021.9413064}

\bibitem{8576421}
Kamrul~Hasan, S.M., Linte, C.A.: A modified u-net convolutional network
  featuring a nearest-neighbor re-sampling-based elastic-transformation for
  brain tissue characterization and segmentation. In: Proc. IEEE WNYISPW.
  pp.~1--5 (2018). \doi{10.1109/WNYIPW.2018.8576421}

\bibitem{ma2021abdomenct}
Ma, J., Zhang, Y., Gu, S., \emph{et al.}: Abdomenct-1k: Is abdominal organ
  segmentation a solved problem. IEEE T PATTERN ANAL  (2021)

\bibitem{cao2020dual}
Cao, H., Liu, H., Song, E., Hung, C.C., Ma, G., Xu, X., Jin, R., Lu, J.:
  Dual-branch residual network for lung nodule segmentation. Applied Soft
  Computing  \textbf{86},  105934 (2020)

\bibitem{isensee2021nnu}
Isensee, F., Jaeger, P.F., Kohl, S.A., \emph{et al.}: nnu-net: a
  self-configuring method for deep learning-based biomedical image
  segmentation. NAT METHODS  \textbf{18}(2),  203--211 (2021)

\bibitem{alom2018recurrent}
Alom, M.Z., Hasan, M., Yakopcic, C., Taha, T.M., Asari, V.K.: Recurrent
  residual convolutional neural network based on u-net (r2u-net) for medical
  image segmentation. arXiv preprint arXiv:1802.06955  (2018)

\bibitem{amorim2019lung}
Amorim, P., Moraes, T.F., Silva, J., \emph{et al.}: Lung nodule segmentation
  based on convolutional neural networks using multi-orientation and patchwise
  mechanisms. In: Proc. VipIMAGE. pp. 286--295 (2019)

\bibitem{tang2019nodulenet}
Tang, H., Zhang, C., Xie, X.: Nodulenet: Decoupled false positive reduction for
  pulmonary nodule detection and segmentation. In: Proc. MICCAI. pp. 266--274.
  Springer (2019)

\bibitem{tolooshams2020channel}
Tolooshams, B., Giri, R., Song, A., \emph{et al.}: Channel-attention dense
  u-net for multichannel speech enhancement. In: Proc. ICASSP. pp. 836--840
  (2020)

\bibitem{zhou2018unet++}
Zhou, Z., Siddiquee, R., Mahfuzur, M., \emph{et al.}: Unet++: A nested u-net
  architecture for medical image segmentation. In: Proc. DLMIA Workshop, pp.
  3--11 (2018)

\bibitem{2020Grad}
Selvaraju, R.R., Cogswell, M., Das, A., \emph{et al.}: Grad-cam: Visual
  explanations from deep networks via gradient-based localization. INT J COMPUT
  VISION  \textbf{128}(2),  336--359 (2020)

\bibitem{ronneberger2015u}
Ronneberger, O., Fischer, P., Brox, T.: U-net: Convolutional networks for
  biomedical image segmentation. In: Proc. MICCAI. pp. 234--241. Springer
  (2015)

\end{thebibliography}

\end{document}